\documentclass[11pt,a4paper]{article}
\usepackage[T1]{fontenc}
\usepackage{lmodern}
\usepackage[a4paper,margin=25mm]{geometry}
\usepackage{amsmath,amssymb,amsthm}
\usepackage{graphicx,booktabs,placeins,needspace,microtype}
\usepackage[round,authoryear]{natbib}
\usepackage[hidelinks]{hyperref}
\newtheorem{theorem}{Theorem}
\newtheorem{proposition}{Proposition}

\AtBeginDocument{\lefthyphenmin=2\relax\righthyphenmin=3\relax}
\newcommand{\1}{\mathbf 1}
\newcommand{\cG}{\mathcal G}
\newcommand{\cD}{\mathcal D}
\newcommand{\lexge}{\succeq_{\rm lex}}
\newcommand{\tea}{\ifmmode\mathit{tea}\else\textit{tea}\fi}
\newcommand{\kea}{\ifmmode\mathit{kea}\else\textit{kea}\fi}
\newcommand{\pr}{\operatorname{pr}}
\newcommand{\Unif}{\operatorname{Unif}}
\newcommand{\paperTitle}{Power improvements and exact shortcuts for closed permutation maxT testing with discrete data}
\newcommand{\alttext}[1]{\par\smallskip{\footnotesize\noindent\textit{Alt text:} #1\par}}
\hypersetup{pdftitle={\paperTitle},pdfauthor={G. Corradini, M. N. Wright, R. Foraita, L. Finos}}
\date{}

\title{\paperTitle}
\author{%
  G. Corradini\textsuperscript{1}, M. N. Wright\textsuperscript{1,2},
  R. Foraita\textsuperscript{1} and L. Finos\textsuperscript{3,4}\\[0.8em]
  \small\textsuperscript{1}Department of Statistical Methods in Epidemiology,\\
  \small Leibniz Institute for Prevention Research and Epidemiology--BIPS\\
  \small Achterstra{\ss}e 30, 28359 Bremen, Germany\\[0.3em]
  \small\textsuperscript{2}Faculty of Mathematics and Computer Science, University of Bremen\\
  \small Bibliothekstra{\ss}e 5, 28359 Bremen, Germany\\[0.3em]
  \small\textsuperscript{3}Department of Statistical Sciences, University of Padova\\
  \small Via C. Battisti 241, 35121 Padova, Italy\\[0.3em]
  \small\textsuperscript{4}Padova Neuroscience Center, University of Padova\\
  \small Via Orus 2/B, 35129 Padova, Italy\\[0.6em]
  \footnotesize\texttt{corradini@leibniz-bips.de}\qquad\texttt{wright@leibniz-bips.de}\\
  \footnotesize\texttt{foraita@leibniz-bips.de}\qquad\texttt{livio.finos@unipd.it}%
}

\begin{document}
\maketitle

\begin{abstract}
In permutation tests with discrete data, especially in small samples, different permutations often yield the same test statistic. The resulting ties can make the test conservative and reduce power. This conservatism can be particularly pronounced for maximum statistics: permutations may have the same maximum but differ at other endpoints, so the remaining endpoint statistics are discarded. We develop two power improvements for closed permutation maxT tests by refining the treatment of ties. The first method uses a single uniform random variable shared across all local tests in the closed testing procedure. This shared-uniform construction maximizes the rejection probability of every elementary hypothesis among all ways of coordinating the same randomized local tests. It also preserves an exact linear step-down shortcut. The second method is the deterministic two-endpoint augmentation (\tea{}) test, which uses the second-largest endpoint statistic to resolve ties in the maximum. This additional information can favour rejection under alternatives while preserving null validity, without auxiliary randomization. Its full closure admits an exact quadratic shortcut. More generally, augmentation using any fixed number of endpoint statistics admits an exact polynomial shortcut. Both procedures retain every rejection made by classical closed maxT procedures and provide strong finite-sample control of the family-wise error rate under intersection-null permutation invariance. In simulations with sparse binary outcomes, both methods improve power over classical closed maxT testing. When many endpoints are tested, the \tea{} test also often outperforms shared-uniform randomization. 
\end{abstract}
\noindent\textit{Some key words:}
Closed testing; Discreteness; Family-wise error rate; Permutation test; Randomization; Tie-breaking.

\section{Introduction}

Exact tests control the type I error rate in finite samples under the assumed null model, but discreteness can make them conservative \citep{tocher1950,lehmannromano2005}, as illustrated by Fisher's lady tasting tea experiment and its few attainable $p$-values \citep{fisher1935}. Counting all ties in permutation tails preserves validity but can reduce power \citep{agresti2013}, also for procedures controlling the family-wise error rate \citep{winkler2016}. Classical closed permutation maxT testing \citep{marcus1976closed,westfallyoung1993} uses the maximum within each intersection and admits an exact step-down shortcut requiring only $m$ local tests for $m$ elementary hypotheses. Conservatism can be particularly pronounced: permutations may share the same maximum while differing at other endpoints, so information in the remaining endpoint statistics is discarded.

One approach to reducing conservatism is the mid-$p$ method, which counts half the probability of a tie \citep{lancaster1961}. However, it need not control the type I error at the nominal level \citep{agresti2013}, so it does not generally provide the local validity required by closure. Randomization instead counts a uniformly random fraction of the tie probability and removes local conservatism while preserving validity \citep{tocher1950,lehmannromano2005}. Although randomized marginal $p$-values have been studied in multiple testing \citep{habigerpena2011,habiger2015}, the coordination of auxiliary randomization across the local tests of closed permutation maxT procedures has not, to the best of our knowledge, been investigated.

Our first method uses a single uniform random variable shared across all local tests. For fixed data and permutations, this construction maximizes the rejection probability of every elementary hypothesis among all ways of coupling the same randomized local tests. It also preserves the exact linear step-down shortcut. Randomized decisions can nevertheless differ for the same data, limiting their practical appeal \citep{agresti2013}. Our second method, two-endpoint augmentation (\tea{}), uses the second-largest endpoint statistic to resolve ties in the maximum without auxiliary randomization. This statistic can distinguish configurations that have the same maximum but are differently informative under an alternative. More generally, $k$-endpoint augmentation (\kea{}) uses the largest $k$ endpoint statistics, with \tea{} corresponding to $k=2$. These constructions use lexicographic comparisons within the framework of ordered test statistics studied by \citet{gurevichvovk2019}.

Both methods retain every classical closed maxT rejection and provide strong finite-sample control of the family-wise error rate under intersection-null permutation invariance. We establish an exact quadratic shortcut for \tea{} closure and an exact polynomial shortcut for each fixed $k$. Under full enumeration, augmentation is deterministic; sampling permutations introduces only the usual Monte Carlo variability. Simulations with sparse binary outcomes illustrate gains over classical closed maxT testing and show that \tea{} can also outperform shared-uniform randomization when many endpoints are tested.

\section{Permutation tests and closed testing}\label{sec:setup}
Let $H_1,\ldots,H_m$ be elementary hypotheses, write $[m]=\{1,\ldots,m\}$, and let $H_I=\bigcap_{j\in I}H_j$ for nonempty $I\subseteq[m]$. Denote the full data by $Z$ and the endpoint data in $I$ by $Z_I$. Under each $H_I$, assume
\begin{equation}\label{eq:invariance}
 Z_I\overset{d}=gZ_I\qquad(g\in\cG),
\end{equation}
where $\mathcal G$ is a finite group of admissible permutations, common to all intersections. This condition must hold when hypotheses outside $I$ are false. In a two-group comparison, for example, it is joint exchangeability of the outcome vectors in $I$ across group positions. Equality of individual marginal distributions alone does not ensure this joint condition. Dependence among endpoints is otherwise allowed.

A test of an intersection hypothesis $H_I$ is called a local test. For endpoint $j$, let $T_j=T_j(Z_j)$ be a real-valued statistic, with larger values more extreme. The statistic is defined for every permuted dataset and depends on no other endpoint. All intersections use the same permutation list $\mathcal P=(g_1,\ldots,g_w)$, with $g_1={\rm id}$. Either $\mathcal P$ enumerates $\cG$, so $w=|\cG|$, or $g_2,\ldots,g_w$ are independent uniform draws with replacement from $\cG$, independent of the data. Permutations are indexed by $r=1,\ldots,w$, with $r=1$ corresponding to the observed data. Repeated draws retain their multiplicity. Set $T_j^{(r)}=T_j(g_rZ_j)$ and
\[
 M_I^{(r)}=\max_{j\in I}T_j^{(r)},\qquad
 a_I=\frac{1}{w}\sum_{r=1}^{w}\1\{M_I^{(r)}>M_I^{(1)}\},\qquad
 b_I=\frac{1}{w}\sum_{r=1}^{w}\1\{M_I^{(r)}=M_I^{(1)}\}.
\]
The identity ensures $b_I>0$. Classical and randomized local $p$-values are
\begin{equation}\label{eq:local}
 p_I^{\mathit{cl}}=a_I+b_I,\qquad p_I^{\mathit{rand}}=a_I+b_IU_I,
 \qquad U_I\sim\Unif(0,1).
\end{equation}

Conditional on $\cD=(Z,\mathcal P)$, a coupling is a joint distribution of $(U_I)_{\emptyset\ne I\subseteq[m]}$ whose one-dimensional marginals are $\Unif(0,1)$. The dependence among the uniforms may depend on $\cD$. Conditionally independent uniforms and a single shared-uniform are two admissible couplings. Under $H_I$, $p_I^{\mathit{cl}}$ is super-uniform and $p_I^{\mathit{rand}}$ is uniform, averaging over the data, auxiliary randomization and any permutation sampling \citep{hemerikgoeman2018}.

For any family of local $p$-values, closed testing gives the adjusted $p$-value
\begin{equation}\label{eq:closure}
 \widetilde p_j=\max_{I\ni j}p_I.
\end{equation}
Thus, $H_j$ is rejected at level $\alpha$ only if every intersection containing $j$ is rejected. If $I_0$ indexes the true elementary hypotheses, any false rejection entails rejection of $H_{I_0}$. Local validity therefore gives strong family-wise error control, meaning control under every configuration of true and false hypotheses. Independence between local tests is unnecessary. Moreover, $p_I^{\mathit{rand}}\leq p_I^{\mathit{cl}}$ for every intersection, so every coupling retains all classical closed-test rejections on the same data and permutation list.

We write $\widetilde p_j^{\mathit{cl}}$ for classical maxT closure, $\widetilde p_j^{\mathit{ind}}$ for randomized maxT closure with conditionally independent intersection uniforms, and $\widetilde p_j^U$ for closure with one shared-uniform $U_I=U$.

\section{Coordinated randomization}\label{sec:shared}
\subsection{Optimal coupling of local tests}
We first consider how to coordinate the auxiliary randomization across local tests.
Conditional on $\cD$, the local rejection probability at level $0\leq\alpha\leq1$ is $\gamma_I(\alpha)=\min\{1,\max\{0,(\alpha-a_I)/b_I\}\}$.
The data and permutation list remain fixed in the next result; only the auxiliary variables vary.

\begin{proposition}\label{prop:coupling}
Among all couplings for which each $U_I$ is uniform conditional on $\cD$, a shared-uniform maximizes the conditional closed-test rejection probability for every elementary hypothesis simultaneously at every significance level. In particular,
\begin{equation}\label{eq:coupling}
 \begin{split}
 \pr_U(\widetilde p_j^{\mathit{ind}}\leq\alpha\mid\cD)
 &=\prod_{I\ni j}\gamma_I(\alpha),\\
 \sup_{\rm couplings}\pr_U(\widetilde p_j\leq\alpha\mid\cD)
 =\pr_U(\widetilde p_j^U\leq\alpha\mid\cD)
 &=\min_{I\ni j}\gamma_I(\alpha).
 \end{split}
\end{equation}
\end{proposition}
\begin{proof}
Closed rejection of $H_j$ is the intersection of its local rejection events. The probability of that intersection is at most the smallest marginal probability. Under the shared-uniform coupling, the events correspond to nested intervals $\{U\leq\gamma_I(\alpha)\}$, whose intersection attains the bound. With independent uniforms, their probabilities multiply.
\end{proof}

Averaging over the data and any permutation sampling shows that shared-uniform randomization maximizes power at each false elementary hypothesis and hence the expected number of true discoveries, among couplings of the local tests in \eqref{eq:local}. When all $L$ local tests have the same conditional rejection probability $\gamma\in(0,1)$, independent uniforms yield a simultaneous rejection probability of $\gamma^L$, whereas a shared-uniform yields $\gamma$. Further details are given in Supplementary Section~S2.1.

\subsection{The linear shortcut}\label{sec:shared-shortcut}
Order the observed statistics as $T_{j_{(1)}}^{(1)}\geq\cdots\geq T_{j_{(m)}}^{(1)}$, breaking ties by a fixed deterministic rule. Here, $j_{(t)}$ is the endpoint at observed position $t$, and $I_t=\{j_{(t)},\ldots,j_{(m)}\}$ is the suffix starting at that position, for $t=1,\ldots,m$. This endpoint order is held fixed across permutations.

\begin{theorem}\label{thm:shared}
Shared-uniform maxT closure satisfies
\begin{equation}\label{eq:shared-shortcut}
 \widetilde p_{j_{(t)}}^U
 =\max_{1\leq\ell\leq t}\{a_{I_\ell}+b_{I_\ell}U\},
 \qquad t=1,\ldots,m.
\end{equation}
Under \eqref{eq:invariance}, the procedure strongly controls the family-wise error rate, averaging over $U$ and any permutation sampling, and satisfies $\widetilde p_j^U\leq\widetilde p_j^{\mathit{cl}}$ pointwise. Given the endpoint statistics, the shortcut requires $O(m\log m)$ operations to sort the observed statistics and $O(wm)$ to evaluate $m$ local tests, for a total of $O(m\log m+wm)$.
\end{theorem}
\begin{proof}
If $I\subseteq J$ and $M_I^{(1)}=M_J^{(1)}$, then $a_I\leq a_J$ and $a_I+b_I\leq a_J+b_J$. For each fixed $u\in[0,1]$,
\[
 a_I+ub_I=(1-u)a_I+u(a_I+b_I)
 \leq(1-u)a_J+u(a_J+b_J)=a_J+ub_J.
\]
For any $I\ni j_{(t)}$, let $\ell$ be its smallest observed position. Then $\ell\leq t$, $I\subseteq I_\ell$, and their observed maxima are equal; hence $p_I^U\leq p_{I_\ell}^U$. Conversely, every suffix $I_\ell$ with $\ell\leq t$ contains $j_{(t)}$. Thus, taking maxima proves \eqref{eq:shared-shortcut}. Local validity and closure ensure error control, while the pointwise local inequalities imply the comparison with classical closure. All suffix maxima can be computed in a single backward pass for each permutation.
\end{proof}

The shortcut reproduces full closure for every fixed value of $U$. Strong family-wise error control holds when $U$ is
drawn uniformly, with probability taken over the data, $U$ and any permutation sampling. Fixing $U$ at a convenient value does not preserve this guarantee. Independent intersection uniforms need not preserve the local inequalities in the proof and therefore do not generally permit the same shortcut. A counterexample and the treatment of tied observed endpoints are given in Supplementary Section~S2.

\section{Endpoint augmentation}\label{sec:tea}
\subsection{The \kea{} test}
We next use additional endpoint statistics to resolve ties without auxiliary randomization. Fix an integer $k\geq1$ in advance and use it for every intersection; define $d_I=\min\{k,|I|\}$. Let $t_{1,I}^{(r)}\geq\cdots\geq t_{|I|,I}^{(r)}$ be the endpoint statistics in $I$. For each permutation $r$, order them decreasingly in $I$, counting ties at distinct endpoints with full multiplicity. The $k$-endpoint augmentation (\kea{}) statistic is
\begin{equation}\label{eq:tea-statistic}
 K_{I,k}^{(r)}=(t_{1,I}^{(r)},\ldots,t_{d_I,I}^{(r)}).
\end{equation}
The endpoints are selected afresh within each intersection and each permutation. Vectors of the same length are compared lexicographically: $x\lexge y$ if $x=y$, or if the first component at which they differ is larger in $x$. Thus the second-largest statistic resolves ties in the maximum, the third resolves remaining ties in the first two components, and so on. For $|I|<k$, all available endpoint statistics are used; for a singleton the comparison is scalar.

The local \kea{} $p$-value and its closed-testing adjustment are
\begin{equation}\label{eq:tea-pvalue}
 p_I^{(k)}=\frac{1}{w}\sum_{r=1}^{w}
 \1\{K_{I,k}^{(r)}\lexge K_{I,k}^{(1)}\},\qquad
 \widetilde p_j^{(k)}=\max_{I\ni j}p_I^{(k)}.
\end{equation}
The case $k=1$ is classical maxT. The case $k=2$ is the two-endpoint augmentation (\tea{}) test, with $p_I^{\tea}=p_I^{(2)}$ and $\widetilde p_j^{\tea}=\widetilde p_j^{(2)}$. For $|I|\geq2$,
\[
 p_I^{\tea}=a_I+\frac{1}{w}\sum_{r=1}^{w}
 \1\{t_{1,I}^{(r)}=t_{1,I}^{(1)},\ t_{2,I}^{(r)}\geq t_{2,I}^{(1)}\}.
\]
Under \eqref{eq:invariance}, $p_I^{(k)}$ is super-uniform under $H_I$ for every fixed $k$, by permutation rank validity \citep{gurevichvovk2019}; see Supplementary Section~S1. Adding a component can only remove permutations tied in all preceding components from the inclusive tail. Consequently, for every $k\geq1$,
\begin{equation}\label{eq:tea-dominance}
 \begin{split}
 p_I^{(k+1)}&\leq p_I^{(k)}\leq p_I^{(1)}=p_I^{\mathit{cl}},\\
 \widetilde p_j^{(k+1)}&\leq\widetilde p_j^{(k)}\leq\widetilde p_j^{(1)}=\widetilde p_j^{\mathit{cl}}.
 \end{split}
\end{equation}
Thus, every fixed $k$ gives strong family-wise error control and power no smaller than classical closure under every alternative, with a strict increase whenever additional rejections have positive probability. Increasing $k$ preserves all rejections obtained with fewer components. Singletons remain classical, and the local statistic stops changing once $k\geq|I|$. Remaining ties are included in the tail.

The \kea{} test, including \tea{}, is deterministic under full enumeration: the data determine its tie-breaking comparisons without an auxiliary random variable. With sampled permutations, it is deterministic for a fixed permutation list and retains only Monte Carlo variability. With $w$ evaluated permutations, its values remain on the grid of multiples of $1/w$. The refinement reduces tail counts rather than changing this grid.

\subsection{Exact shortcuts}
Use the observed endpoint order and suffixes from Section~\ref{sec:shared-shortcut}. For $1\leq k\leq m$, let $\boldsymbol{\ell}=(\ell_1,\ldots,\ell_k)$ be a tuple of observed positions with $1\leq\ell_1<\cdots<\ell_k\leq m$. Define its completion and the retained family by
\begin{equation}\label{eq:quadratic-family}
 \begin{split}
 C_{\boldsymbol{\ell}}&=\{j_{(\ell_1)},\ldots,j_{(\ell_{k-1})}\}\cup I_{\ell_k},\\
 \mathcal F_k&=\{C\subseteq[m]:1\leq|C|<k\}
 \ \cup\ \{C_{\boldsymbol{\ell}}:1\leq \ell_1<\cdots<\ell_k\leq m\}.
 \end{split}
\end{equation}
The completion $C_{\boldsymbol{\ell}}$ contains the $k-1$ explicitly listed endpoints and the suffix starting at position $\ell_k$. For $k=1$, the separately listed endpoints and the first family are empty, so $\mathcal F_1$ consists solely of suffixes.

\begin{theorem}\label{thm:tea}
For each fixed $1\leq k\leq m$, full closure of the local tests \eqref{eq:tea-pvalue} satisfies
\begin{equation}\label{eq:tea-shortcut}
 \widetilde p_j^{(k)}=\max_{C\in\mathcal F_k:\,j\in C}p_C^{(k)}.
\end{equation}
Under \eqref{eq:invariance}, it controls family-wise error strongly and satisfies \eqref{eq:tea-dominance}. The shortcut requires $N_k(m)=\sum_{s=1}^k\binom ms$ local tests and, after ordering and computing the endpoint statistics, $O\{wkN_k(m)\}$ operations. For fixed $k$, the observed sorting  $O(m\log m)$ cost gives an overall complexity of $O(m\log m+wm^k)$. Thus,  for \tea{} the complexity is $O(m\log m+wm^2)$, excluding permutation generation and endpoint-statistic evaluation.
\end{theorem}
\begin{proof}
Consider an intersection $I$ of size at least $k$, whose first $k$ observed positions are $\ell_1<\cdots<\ell_k$. Its completion $C_{\boldsymbol{\ell}}$ contains $I$ and has the same $k$ largest observed statistics. Adding endpoints cannot decrease any of the first $k$ order statistics in any permutation. Hence $K_{I,k}^{(1)}=K_{C_{\boldsymbol{\ell}},k}^{(1)}$ and $K_{I,k}^{(r)}\preceq_{\rm lex}K_{C_{\boldsymbol{\ell}},k}^{(r)}$, which implies $p_I^{(k)}\leq p_{C_{\boldsymbol{\ell}}}^{(k)}$. Intersections of size less than $k$ are retained separately. Every omitted local $p$-value is therefore bounded by a retained value for a superset, proving \eqref{eq:tea-shortcut}. The completions have distinct first $k$ observed positions and size at least $k$, giving the stated count. Rank validity and \eqref{eq:tea-dominance} give error control and dominance. For computation, store the largest $k$ available statistics of each suffix in each permutation and merge them with the sorted statistics of the $k-1$ separately retained endpoints. Recursive construction of these sorted lists and aggregation of the local $p$-values give the stated cost; an algorithm and aggregation details are given in Supplementary Section~S3.2.
\end{proof}

For \tea{}, write $C_{\ell_1,\ell_2}=\{j_{(\ell_1)}\}\cup I_{\ell_2}$, where $1\leq\ell_1<\ell_2\leq m$. Then $N_2(m)=m+\binom m2=m(m+1)/2$, recovering the exact quadratic shortcut. For $k=3$, retaining all singletons, pairs and completions $C_{\ell_1,\ell_2,\ell_3}=\{j_{(\ell_1)},j_{(\ell_2)}\}\cup I_{\ell_3}$, with $1\leq\ell_1<\ell_2<\ell_3\leq m$, gives a cubic shortcut. More generally, the cost has degree $k$ in $m$ for fixed $k$; when $k=m$, the family contains all $2^m-1$ nonempty intersections.

For four endpoints ordered as $1,2,3,4$, the retained nonsingleton sets for \tea{} are $\{1,2,3,4\}$, $\{1,3,4\}$, $\{1,4\}$, $\{2,3,4\}$, $\{2,4\}$ and $\{3,4\}$. For example, $\{1,3\}$ and $\{1,3,4\}$ have the same observed pair, $(T_1^{(1)},T_3^{(1)})$. In each permutation, adding endpoint 4 cannot decrease either of the two largest statistics. Since both pairs are compared with the same observed pair, every permutation counted in the tail for $\{1,3\}$ is also counted for
$\{1,3,4\}$, giving $p_{\{1,3\}}^{\tea}\leq p_{\{1,3,4\}}^{\tea}$.

We focus on the \tea{} test as a practical compromise between power and computational efficiency: it uses the first augmentation beyond the maximum while retaining a quadratic shortcut. Increasing $k$ can yield further gains when many endpoints are tested, at the cost of a higher-order shortcut. Supplementary Section~S5.5 compares $k=2,3,4$ under the same simulation design.

\subsection{Power comparisons within ties}\label{sec:tea-power}
The \kea{} test can gain power over randomized maxT because datasets tied at the maximum need not be equally informative under an alternative. To make this precise, fix $k\geq2$ and an intersection $I$ with at least two endpoints, and condition on its fully enumerated permutation orbit $\mathcal O_I=\{gZ_I:g\in\cG\}$. For an attainable maximum $c$, let $C=\{z\in\mathcal O_I:\max_{j\in I}T_j(z_j)=c\}$ be its tie class. Let $P_{0,C}$ and $P_{1,C}$ be the conditional null and alternative laws on $C$, with likelihood ratio $\ell_C=dP_{1,C}/dP_{0,C}$. The orbit is held fixed throughout, and the alternative is assumed to assign positive probability to $C$.

Write $a$ and $b$ for the null proportions on the orbit with maxima strictly greater than and equal to $c$, respectively. At level $\alpha$, randomized maxT rejects each configuration in $C$ with probability $\gamma=\min\{1,\max\{0,(\alpha-a)/b\}\}$. For $z\in C$, let $S_I(z)=(t_{2,I}(z),\ldots,t_{d_I,I}(z))$ be the remaining ordered vector and define its conditional tail rank as $R_C(z)=P_{0,C}\{S_I(Z')\lexge S_I(z)\}$, where $Z'\sim P_{0,C}$.
The local augmentation $p$-value is then $p_I^{(k)}(z)=a+bR_C(z)$. Thus augmentation replaces the auxiliary uniform by a rank determined by the data within the maximum tie class. Under $P_{0,C}$ this rank is super-uniform: $P_{0,C}\{R_C(Z)\leq u\}\leq u$ for $0\leq u\leq1$. If $S_I$ distinguishes all $|C|$ configurations, the rank is uniform on $\{1/|C|,\ldots,1\}$; otherwise, residual ties can make the rank distribution more conservative.

Let $F_{1,C}(u)=P_{1,C}\{R_C(Z)\leq u\}$. The conditional powers of augmentation and randomized maxT are $F_{1,C}(\gamma)$ and $\gamma$, respectively. To connect this rank comparison to the information within the tie, set $r_\alpha(z)=\1\{R_C(z)\leq\gamma\}$ and $q_\alpha=E_0(r_\alpha\mid C)\leq\gamma$. Since $E_0(\ell_C\mid C)=1$,
\begin{equation}\label{eq:tie-power}
 F_{1,C}(\gamma)-\gamma
 =\operatorname{cov}_0(r_\alpha,\ell_C\mid C)-(\gamma-q_\alpha).
\end{equation}
The covariance measures the information used by the deterministic rejection rule; $\gamma-q_\alpha$ is the remaining loss from discreteness. Augmentation is more powerful within $C$ precisely when $F_{1,C}(\gamma)>\gamma$, or equivalently when the covariance exceeds this loss. Lexicographically larger remaining vectors can supply this information if they identify configurations more likely under the alternative. For \tea{}, this concerns larger values of the second-largest endpoint statistic. At equal conditional size, a positive covariance alone gives a strict power gain. Supplementary Section~S4.3 gives the derivation and a sufficient monotonicity condition.

Shared-uniform randomization and every fixed-$k$ augmentation with $k\geq2$ retain all classical closed-test rejections, giving no smaller power at any false elementary hypothesis. There is, however, no general power ordering between the two approaches. The comparison above concerns local tests, whose gains need not survive closure. Supplementary Section~S4 gives a finite experiment in which \tea{} closure has greater family rejection power than shared-uniform closure. In that experiment, its family rejection event is uniformly most powerful for the specified conditional complete-null testing problem.

\section{Simulations}\label{sec:simulations}
We compare the procedures for independent binary endpoints in two balanced groups, with total sample size $n\in\{20,50,100,250\}$ and family size $m\in\{10,100,1000\}$. The control event probability is 0.005. In every dataset, a randomly selected 10\% of endpoints have an increased treatment probability; the remaining 90\% are true nulls. Risk differences are calibrated across $n$ and $m$ as specified in Supplementary Section~S5, so that comparisons across sample sizes do not represent a fixed alternative. The statistic is the one-sided pooled standardized difference in proportions. Every scenario uses 1000 datasets and 1000 sampled permutations plus the identity, with the same data and permutation lists for the permutation procedures. Power over the signal endpoints and family-wise error over the true nulls are estimated from these same datasets, while all $m$ endpoints enter each procedure. One auxiliary uniform is drawn per dataset and shared across intersections. The nominal level is 0.05.

\begin{figure}[t]
\centering
\includegraphics[width=\linewidth]{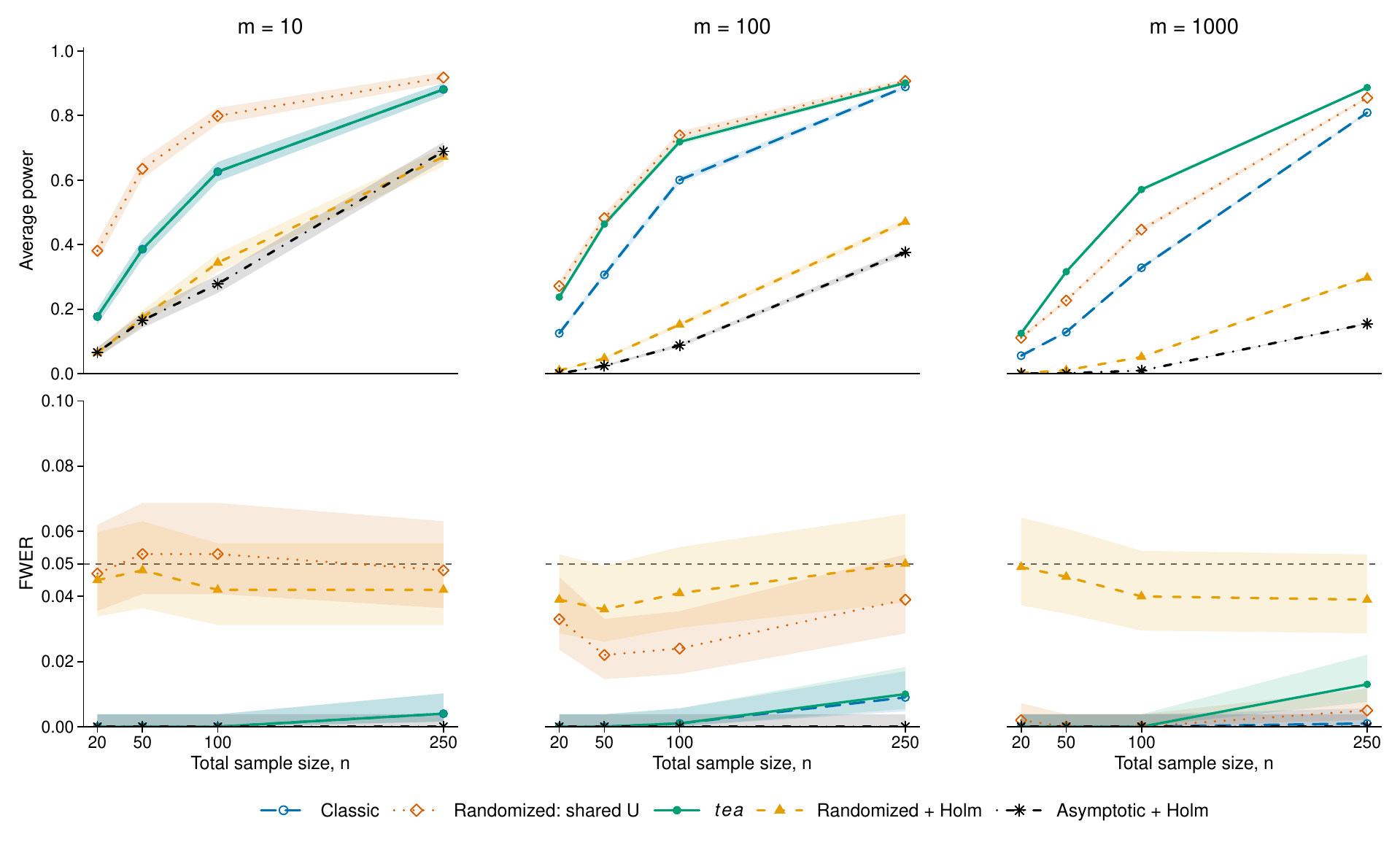}
\caption{Average power over signals (top) and family-wise error rate over true nulls (bottom) for independent binary endpoints with 10\% signals. Bands show 95\% Monte Carlo intervals; dashed lines mark 0.05.}
\label{fig:power-fwer}
\alttext{Classical closed maxT exceeds both Holm comparators. Shared-uniform randomization improves on classical maxT throughout. Two-endpoint augmentation matches classical power with 10 endpoints, improves substantially with 100 endpoints, and exceeds shared-uniform randomization with 1000 endpoints. Family-wise error is measured among the 90\% true nulls. The augmentation and classical procedures are conservative; shared-uniform randomization is near nominal with 10 endpoints and more conservative with larger families.}
\end{figure}

Figure~\ref{fig:power-fwer} compares classical and shared-uniform closed maxT, \tea{} closure, and two Holm procedures \citep{holm1979}: randomized marginal $p$-values with independent endpoint uniforms, and asymptotic normal marginal $p$-values. Classical and shared-uniform closed maxT are more powerful than these Holm comparators in the displayed scenarios, and sharing the uniform increases power beyond classical maxT.

With 10 endpoints, \tea{} and classical maxT have the same estimated power, while shared-uniform randomization improves on both. With 100 endpoints, \tea{} substantially improves on classical maxT but remains slightly below shared-uniform randomization. With 1000 endpoints, \tea{} exceeds shared-uniform randomization at every displayed sample size. For example, at $n=100$ their powers are 57.1\% and 44.7\%, respectively, compared with 32.8\% for classical maxT. In these alternatives the number of signals grows with the family size, providing more opportunities for the second-largest statistic to distinguish datasets tied at the maximum. This interpretation is specific to the simulated alternatives; a larger family alone does not guarantee a gain.

The \tea{} test achieves these gains while its estimated family-wise error under the partial-null configurations remains at most 0.013. Shared-uniform error rates are near nominal with 10 endpoints and become more conservative with larger families. As discussed in Section~\ref{sec:tea-power}, information within ties can increase power despite conservatism. Strong control under arbitrary null configurations follows from the theory.

\FloatBarrier
\section{Discussion}
The two procedures increase power by coordinating randomized local tests or using additional endpoint information within ties. Shared-uniform randomization is optimal among couplings of the same local tests and preserves the linear shortcut. The \tea{} test provides a deterministic alternative: it replaces an auxiliary random decision by a comparison of the second-largest endpoint statistics, while preserving strong family-wise error control in finite samples and admitting an exact quadratic shortcut. This distinction is practically relevant when conclusions determined partly by an auxiliary random number are undesirable. A power gain from augmentation does not contradict Proposition~\ref{prop:coupling}: that result fixes the randomized local tests and requires uniform auxiliary variables conditional on the full data and permutation list. Under full enumeration, the augmentation ranks are determined by those data and calibrated under the null conditional on the orbit and maximum tie class.

Randomized maxT assigns the same rejection probability to each configuration in a maximum tie class. Endpoint augmentation uses additional evidence whose conditional tail rank may shift towards smaller values under alternatives. Its power advantage depends on whether this information outweighs residual discreteness, as quantified in \eqref{eq:tie-power}.

\Needspace{2\baselineskip}
The shared-uniform construction and its coupling optimality are not restricted to maximum statistics. More generally, let $A_I=A_I(Z_I)$ and $B_I=B_I(Z_I)$ be fixed real-valued statistics. Under \eqref{eq:invariance}, lexicographic permutation testing of $(A_I,B_I)$ preserves finite-sample validity and retains every rejection of the corresponding classical closure based on $A_I$ alone. Thus the augmentation principle underlying the \tea{} test extends to other intersection statistics. The shortcut arguments can likewise extend to other aggregation rules sharing the structural properties used here.

\section*{Supplementary material}
The supplement gives rank proofs and conditioning details, additional results for shared-uniform randomization, the general \kea{} shortcut and its quadratic \tea{} implementation, the finite conditional experiment and the simulation specifications, including experiments with correlated endpoints and higher-order augmentation.

\section*{Data and code availability}
The R/C++ implementation of the proposed tests and the scripts and materials used for the simulation studies are publicly available at \url{https://github.com/bips-hb/tea-maxT-paper}.

\section*{Declaration of the use of generative AI and AI-assisted technologies}
ChatGPT (OpenAI) was used to assist with selected parts of the R/C++ code. The authors reviewed and edited the resulting code as necessary and take full responsibility for the code and the results reported in this manuscript.

\end{document}

\maketitle

\section{Permutation ranks and finite-sample validity}\label{sec:ranks}
\subsection{Notation and assumptions}
The elementary hypotheses are $H_j$, indexed by $[m]=\{1,\ldots,m\}$, and $H_I=\bigcap_{j\in I}H_j$. Under every $H_I$, the corresponding endpoint data satisfy $Z_I\overset{d}=gZ_I$ for each element of a common finite group $\cG$. A test of an intersection hypothesis $H_I$ is called a local test. The statistic $T_j$ uses only endpoint $j$ and fixed design quantities. All intersections use the same permutation list $\mathcal P=(g_1,\ldots,g_w)$, with $g_1={\rm id}$, either enumerating $\cG$, so $w=|\cG|$, or consisting of the identity and $w-1$ independent uniform draws $g_2,\ldots,g_w$ with replacement from $\cG$, independent of the data. Permutations are indexed by $r=1,\ldots,w$, with $r=1$ corresponding to the observed data. Repeated draws retain their multiplicity. We write $T_j^{(r)}=T_j(g_rZ_j)$ and $\cD=(Z,\mathcal P)$.

For $M_I^{(r)}=\max_{j\in I}T_j^{(r)}$, let $a_I$ and $b_I$ be the proportions strictly greater than and equal to $M_I^{(1)}$, respectively. Classical local $p$-values are $p_I^{\mathit{cl}}=a_I+b_I$. Randomized local $p$-values are $p_I^{\mathit{rand}}=a_I+b_IU_I$. A coupling is a joint conditional distribution of $(U_I)_{\emptyset\ne I\subseteq[m]}$ given $\cD$, with $\Unif(0,1)$ marginals; its dependence structure may depend on $\cD$. A shared uniform gives $p_I^U=a_I+b_IU$; conditionally independent intersection uniforms give the independent construction. The notation $\widetilde p_j^{\mathit{cl}}$, $\widetilde p_j^U$, $\widetilde p_j^{\mathit{ind}}$ and $\widetilde p_j^{\tea}$ denotes the corresponding adjusted $p$-values, obtained by maximizing local $p$-values over all intersections containing $j$.

For an integer $k\geq1$ fixed in advance and common to all intersections, the $k$-endpoint augmentation (\kea{}) test uses
\[
 d_I=\min\{k,|I|\},\qquad
 K_{I,k}^{(r)}=(t_{1,I}^{(r)},\ldots,t_{d_I,I}^{(r)}),\qquad
 p_I^{(k)}=\frac{1}{w}\sum_{r=1}^{w}\1\{K_{I,k}^{(r)}\lexge K_{I,k}^{(1)}\},
\]
where $t_{1,I}^{(r)}\geq\cdots\geq t_{|I|,I}^{(r)}$ orders the endpoint statistics in that intersection and permutation, counting equal values from distinct endpoints separately. In lexicographic order, the first unequal component determines the comparison; equal vectors are tied. Thus each successive endpoint statistic resolves ties left by the preceding components. Comparisons are always within the same intersection, so vectors of different lengths are never compared. Intersections with fewer than $k$ endpoints use all available statistics, and singletons use ordinary scalar comparison. The adjusted $p$-value is $\widetilde p_j^{(k)}=\max_{I\ni j}p_I^{(k)}$.

The case $k=1$ gives $p_I^{(1)}=p_I^{\mathit{cl}}$. Two-endpoint augmentation (\tea{}) is the case $k=2$, with $p_I^{\tea}=p_I^{(2)}$ and $\widetilde p_j^{\tea}=\widetilde p_j^{(2)}$. For this case, write $K_I^{(r)}=K_{I,2}^{(r)}$, identifying its one-component vector with a scalar for singletons. Every fixed-$k$ rule is deterministic under full enumeration and introduces no auxiliary randomization; sampling permutations retains only the usual Monte Carlo variability.

\subsection{Finite exchangeable ranks}
Our use of ordered test statistics follows \citet{gurevichvovk2019}. For the finite arguments here, a total preorder $\succeq$ suffices: it is reflexive and transitive and makes any two realised values comparable, while allowing equivalences. Its strict part is denoted by $\succ$, and $x\sim y$ means that both $x\succeq y$ and $y\succeq x$ hold.

\Needspace{14\baselineskip}
\begin{lemma}[Exchangeable ordered ranks]\label{lem:ranks}
Let $X_1,\ldots,X_w$ be exchangeable random elements with a total preorder. Then
\[
 p=\frac{1}{w}\sum_{r=1}^{w}\1\{X_r\succeq X_1\}
\]
satisfies $\pr(p\leq\alpha)\leq\alpha$ for $0\leq\alpha\leq1$. If $U$ is independent and uniform on $(0,1)$, then
\[
 p^{\mathit{rand}}=\frac{1}{w}\sum_{r=1}^{w}\1\{X_r\succ X_1\}
 +\frac{U}{w}\sum_{r=1}^{w}\1\{X_r\sim X_1\}
\]
is uniform on $(0,1)$.
\end{lemma}
\begin{proof}
Condition on the unordered multiset of the $w$ realised values. Exchangeability makes label 1 uniform over its $w$ positions. Order equivalence classes from most to least extreme, breaking remaining equivalences arbitrarily to obtain a linear extension. If $X_1$ occupies position $h$, at least $h$ realised values are at least as extreme as $X_1$, so $p\geq h/w$. At most $\lfloor\alpha w\rfloor$ labels can give $p\leq\alpha$.

For randomization, a class of $c$ equal values with $d$ strictly more extreme entries is selected with probability $c/w$. Conditional on that class, the randomized value is uniform on $[d/w,(d+c)/w]$. The intervals associated with the ordered classes partition $[0,1]$ and have lengths equal to their selection probabilities. Their mixture is therefore uniform. Averaging the conditional statements proves both claims.
\end{proof}

For full enumeration, condition on the orbit $\mathcal O_I=\{gZ_I:g\in\cG\}$ under $H_I$. Null invariance makes the observed configuration uniform on that finite orbit. Every orbit point has the same number of preimages in $\cG$, so counting group elements or counting distinct orbit points gives the same tail proportions. The observed statistic consequently selects each tie class with probability equal to its proportion in the full permutation list. The counting and interval arguments of Lemma~\ref{lem:ranks} apply directly; no exchangeability of a particular fixed ordering of the entire group is needed.

For independently sampled uniform group elements with the identity included, conditional on $\mathcal O_I$, the configurations $Z_I,g_2Z_I,\ldots,g_wZ_I$ are independent uniform draws from that orbit. Indeed, for any orbit point $z$, a uniform group element sends $z$ to every orbit point with equal probability, independently of $z$. The resulting statistics are exchangeable, so Lemma~\ref{lem:ranks} applies. This supplies the required finite-sample rank validity for scalar maxima and the lexicographically ordered vectors used by \kea{}, including the pairs used by \tea{}; see also \citet{hemerikgoeman2018}. Under sampled permutations, the guarantee averages over their sampling. It is not a claim of validity conditional on an arbitrary fixed permutation list.

\subsection{Closure and pointwise improvement}
Let $I_0$ be the set of true elementary hypotheses. If $I_0$ is empty, the family-wise error probability is zero. Otherwise, rejection of any $H_j$ with $j\in I_0$ requires rejection of $H_{I_0}$, since its local $p$-value occurs in the closed maximum for $j$. Thus
\[
 \pr\{\exists j\in I_0:\widetilde p_j\leq\alpha\}
 \leq\pr(p_{I_0}\leq\alpha)\leq\alpha.
\]
This is the closed-testing argument of \citet{marcus1976closed}. The invariance assumption is required under $H_{I_0}$ even when other hypotheses are false. Marginal equality at each endpoint, without the corresponding joint invariance, is not sufficient.

Every randomized local $p$-value is at most $a_I+b_I$, pointwise in the data, permutation list and auxiliary draws. For \kea{}, the first component of $K_{I,k}$ is $M_I$, so
\[
 \{K_{I,k}^{(r)}\lexge K_{I,k}^{(1)}\}
 \subseteq\{M_I^{(r)}\geq M_I^{(1)}\}.
\]
For singletons the two events are identical. More generally, when $k<|I|$, a comparison decided by one of the first $k$ components is unchanged upon adding component $k+1$; only a tie in all first $k$ components can be removed from the inclusive tail. When $k\geq|I|$, the statistic is unchanged. Hence
\[
 p_I^{(k+1)}\leq p_I^{(k)}\leq p_I^{(1)}=p_I^{\mathit{cl}},\qquad
 \widetilde p_j^{(k+1)}\leq\widetilde p_j^{(k)}\leq\widetilde p_j^{\mathit{cl}}.
\]
The inequalities for the adjusted $p$-values follow by maximizing over intersections. The same maximization proves randomized dominance. Thus shared-uniform randomization and every fixed-$k$ augmentation retain all classical rejections at every nominal level on the same data and permutation list, and increasing $k$ retains all rejections from the preceding augmentation. These deterministic comparisons give no smaller power under every alternative, with strict improvement whenever the additional rejection event has positive probability. They do not require the distributional assumptions used to establish null validity. Together with the rank argument and closure, they establish strong family-wise error control and pointwise dominance for \kea{}, with \tea{} as the case $k=2$.

\subsection{Conditioning and attainable levels}
Randomized validity averages over the auxiliary variable. For example, if $X\sim{\rm Bernoulli}(1/2)$, its upper-tail randomized value is $U/2$ when $X=1$ and $(1+U)/2$ when $X=0$. Fixing $U=0.01$ makes its null rejection probability at level 0.05 equal to $1/2$. Recording a draw or its seed permits reproduction of the procedure, but does not make validity conditional on that draw. Replacing $U$ by $1/2$ yields the mid-$p$ value, which need not be valid.

Under full enumeration, let $C_1,\ldots,C_L$ be the tie classes of an ordered statistic on a permutation orbit, from most to least extreme, counted with group multiplicity. Its attainable inclusive $p$-values are
\[
 \alpha_k=\frac{\sum_{\ell=1}^k|C_\ell|}{|\cG|},
 \qquad k=1,\ldots,L,
\]
and $\pr_0(p\leq\alpha_k\mid\mathcal O_I)=\alpha_k$. Between attainable levels the test is conservative. If a primary tie of size $c$ is separated into ordered classes of sizes $c_1,\ldots,c_s$, and $d$ values were strictly more extreme, the refinement introduces cumulative levels $(d+c_1+\cdots+c_h)/|\cG|$, for $h=1,\ldots,s$. It preserves all previously attainable levels. This describes how successive endpoint statistics can reduce discreteness without auxiliary randomization: each added component refines the tie classes left by the preceding components.

For sampled permutations, the inclusive numerical grid remains the multiples of $1/w$. Refinement can decrease the tail count but cannot change that grid, and the identity still ensures a strictly positive inclusive value. Randomized values can lie below $1/w$ because the identity's tied contribution is itself multiplied by $U$. Lemma~\ref{lem:ranks} explains why this is valid for the randomized construction.

\section{Shared-uniform closure}\label{sec:shared}
\subsection{Coupling optimality and its scope}
Condition on $\cD$ and set $\gamma_I(\alpha)=\min\{1,\max\{0,(\alpha-a_I)/b_I\}\}$. Up to events of auxiliary probability zero, local rejection is $\{U_I\leq\gamma_I(\alpha)\}$. For any coupling with conditionally uniform marginals,
\[
 \pr_U\!\left(\bigcap_{I\ni j}\{U_I\leq\gamma_I(\alpha)\}\,\middle|\,\cD\right)
 \leq\min_{I\ni j}\gamma_I(\alpha).
\]
With $U_I=U$ for every $I$, the intersection is $\{U\leq\min_{I\ni j}\gamma_I(\alpha)\}$ and attains the bound. With conditionally independent uniforms its probability is $\prod_{I\ni j}\gamma_I(\alpha)$. This proves Proposition~1 of the main paper, including simultaneous attainment for all endpoints and levels by one coupling.

For any fixed distribution of the data, integration preserves the endpoint rejection inequalities. If $I_1$ is the set of false elementary hypotheses, then
\[
 E\!\left\{\sum_{j\in I_1}\1(\widetilde p_j\leq\alpha)\right\}
 =\sum_{j\in I_1}\pr(\widetilde p_j\leq\alpha)
\]
is also maximized by the shared uniform. The comparison fixes the local quantities $a_I,b_I$ and the uniform marginals. It gives neither a pointwise ordering between different realised randomizations nor an ordering over all possible local tests. In particular, the augmentation rank $R_C$ in Section~\ref{sec:tie-information} is determined by the data, rather than uniform conditional on $\cD$. Its null calibration instead conditions on the orbit and maximum tie class. A power gain from augmentation therefore does not contradict the coupling result.

Independent randomization can lose power under closure despite retaining every classical rejection. If $L$ relevant intersections have the same $a,b$, their largest local $p$-value is $a+b\max_{1\leq h\leq L}U_h$. For $0\leq v\leq1$, $\pr(\max_h U_h\leq v)=v^L$, so this maximum converges in probability to $a+b$. At a level between $a$ and $a+b$, the contribution of these intersections to joint rejection is $\{(\alpha-a)/b\}^L$ instead of $(\alpha-a)/b$.

\subsection{Proof of the suffix identity}
Let $j_{(1)},\ldots,j_{(m)}$ order the observed statistics decreasingly, with deterministic resolution of ties, and let $I_t=\{j_{(t)},\ldots,j_{(m)}\}$ for $t=1,\ldots,m$. If $I\subseteq J$ and $M_I^{(1)}=M_J^{(1)}$, monotonicity of maxima gives
\[
 a_I\leq a_J,\qquad a_I+b_I\leq a_J+b_J.
\]
The equality of observed maxima is essential. For fixed $u\in[0,1]$, the local $p$-value is a convex combination of these strict and inclusive tails:
\[
 p_I^u=(1-u)a_I+u(a_I+b_I)
 \leq(1-u)a_J+u(a_J+b_J)=p_J^u.
\]
Now fix $I\ni j_{(t)}$ and let $\ell$ be the smallest observed position in $I$. Then $\ell\leq t$, $I\subseteq I_\ell$ and the observed maxima agree. Consequently, $p_I^U\leq p_{I_\ell}^U$. Conversely, each suffix $I_\ell$ with $\ell\leq t$ contains $j_{(t)}$ and already occurs in its full closed maximum. Hence
\[
 \widetilde p_{j_{(t)}}^U=\max_{I\ni j_{(t)}}p_I^U
 =\max_{1\leq\ell\leq t}p_{I_\ell}^U.
\]
Together with Section~\ref{sec:ranks}, this proves the validity and dominance claims of Theorem~1 of the main paper. The identity holds for every evaluated statistic matrix and every fixed $u$, whereas finite-sample error control uses the null invariance and a properly generated uniform.

\subsection{Independent uniforms and observed ties}
Independent uniforms need not preserve the required nesting. If all endpoint statistics are constant on the orbit, then $a_I=0$, $b_I=1$ and $p_I^{\mathit{rand}}=U_I$ for every intersection. With two endpoints ordered as 1 then 2, full independent closure gives
\[
 \widetilde p_1^{\mathit{ind}}=\max\{U_{\{1\}},U_{\{1,2\}}\},
\]
whereas the suffix calculation gives only $U_{\{1,2\}}$. These differ whenever $U_{\{1\}}>U_{\{1,2\}}$, an event of positive probability. Full closure remains valid, but the usual shortcut is not its exact evaluation. Under a shared uniform all adjusted $p$-values instead equal $U$, so the entire constant family is rejected with probability $\alpha$. 

If $T_j^{(1)}=T_k^{(1)}$, shared-uniform scalar maxT closure gives $\widetilde p_j^U=\widetilde p_k^U$. For every $I\ni j$, put $J=I\cup\{k\}$. The observed maxima agree, so $p_I^U\leq p_J^U\leq\widetilde p_k^U$. Maximizing over $I$ gives $\widetilde p_j^U\leq\widetilde p_k^U$, and symmetry gives equality. This holds even if the endpoint permutation distributions differ and makes the arbitrary deterministic ordering within an observed tie immaterial to the adjusted $p$-values.

\subsection{Linear computation}
After computing the observed order, each permutation row supplies all suffix maxima in a backward pass:
\[
 L_m^{(r)}=T_{j_{(m)}}^{(r)},\qquad
 L_t^{(r)}=\max\{T_{j_{(t)}}^{(r)},L_{t+1}^{(r)}\}\quad(t=m-1,\ldots,1).
\]
Accumulate the numbers $G_t$ and $E_t$ of strict exceedances and ties of $L_t^{(r)}$ relative to $L_t^{(1)}=T_{j_{(t)}}^{(1)}$, including the identity. Then $p_{I_t}^U=(G_t+UE_t)/w$ and $p_{I_t}^{\mathit{cl}}=(G_t+E_t)/w$. A forward cumulative maximum gives their respective adjusted $p$-values. The closure calculation takes $O(wm)$ time and can use $O(m)$ additional working memory when permutation rows are processed sequentially. Including the $O(m\log m)$ observed sorting cost gives $O(m\log m+wm)$, separately from generating permutations and evaluating endpoint statistics. 

\section{Exact shortcuts for endpoint augmentation}\label{sec:quadratic}
\subsection{Completion of intersections}
Use the observed order from Section~\ref{sec:shared}. For $1\leq k\leq m$ and $\boldsymbol{\ell}=(\ell_1,\ldots,\ell_k)$ with $1\leq \ell_1<\cdots<\ell_k\leq m$, define
\[
 \begin{split}
 C_{\boldsymbol{\ell}}&=\{j_{(\ell_1)},\ldots,j_{(\ell_{k-1})}\}\cup I_{\ell_k},\\
 \mathcal F_k&=\{C\subseteq[m]:1\leq|C|<k\}\cup\{C_{\boldsymbol{\ell}}:1\leq\ell_1<\cdots<\ell_k\leq m\}.
 \end{split}
\]
For $k=1$, this is the suffix family. If $I\subseteq J$ and $|I|\geq k$, adding endpoints cannot decrease any of the first $k$ decreasing order statistics. Therefore $K_{I,k}^{(r)}\preceq_{\rm lex}K_{J,k}^{(r)}$ in every permutation. For an intersection $I$ whose first $k$ observed positions are $\ell_1<\cdots<\ell_k$, all other members have position greater than $\ell_k$, so $I\subseteq C_{\boldsymbol{\ell}}$. All added endpoints have observed statistics no greater than that at position $\ell_k$. Thus
\[
 K_{I,k}^{(1)}=K_{C_{\boldsymbol{\ell}},k}^{(1)},\qquad
 \1\{K_{I,k}^{(r)}\lexge K_{I,k}^{(1)}\}
 \leq\1\{K_{C_{\boldsymbol{\ell}},k}^{(r)}\lexge K_{C_{\boldsymbol{\ell}},k}^{(1)}\}.
\]
It follows that $p_I^{(k)}\leq p_{C_{\boldsymbol{\ell}}}^{(k)}$. All intersections of size less than $k$ are retained separately. Every omitted local $p$-value is bounded by a retained local $p$-value for a superset containing the same elementary hypotheses. Since every retained set is itself an intersection in the full closure,
\[
 \widetilde p_j^{(k)}=\max_{C\in\mathcal F_k:\,j\in C}p_C^{(k)}.
\]
Distinct increasing $k$-tuples give distinct completions, since their entries are exactly the first $k$ observed positions of the completed set. Every completion has at least $k$ members, so none duplicates a separately retained smaller intersection. The number of retained sets is therefore
\[
 N_k(m)=\sum_{s=1}^{k-1}\binom ms+\binom mk=\sum_{s=1}^k\binom ms.
\]
This proves the closure identity and count in Theorem~2 of the main paper, with validity and dominance supplied by Section~\ref{sec:ranks}. The proof also covers ties: equal endpoint values retain their multiplicities, and any deterministic ordering of tied observed endpoints yields the same full closure. For $k=m$, every nonempty intersection is retained. Values $k>m$ give the same tests as $k=m$ and require no further components.

\subsection{Computation and aggregation for general \kea{}}\label{sec:kea-computation}
The following algorithm evaluates the retained local tests for
a fixed $1\leq k\leq m$, given the endpoint statistics
$T_j^{(r)}$.

\begin{enumerate}
\item
Order the endpoints by their observed statistics, using the fixed
tie-breaking rule, and form the family $\mathcal F_k$.
Initialize an inclusive tail count $B_C=0$ for every $C\in\mathcal F_k$.

\item
Process permutations in the order $r=1,\ldots,w$, starting with the observed row.
For each row, compute the largest $\min\{k,|I_t|\}$ statistics
of every suffix $I_t$, $t=1,\ldots,m$, by a backward pass, retaining multiplicities.
Also construct sorted statistic lists for subsets of fewer than
$k$ observed positions, recursively inserting one additional statistic.

For a retained set of size less than $k$, its sorted list is
$K_{C,k}^{(r)}$. For a completion $C_{\boldsymbol{\ell}}$, merge
the sorted statistics of its $k-1$ separately retained endpoints
with the stored list for $I_{\ell_k}$, keeping the largest $k$
values. Store $K_{C,k}^{(1)}$ when processing the observed row.
For every row, update
\[
 B_C\leftarrow B_C+
 \1\{K_{C,k}^{(r)}\lexge K_{C,k}^{(1)}\}.
\]
For $k=1$, the separately retained list is empty and only
suffix maxima are needed.

\item
Set $p_C^{(k)}=B_C/w$ and compute the adjusted $p$-values using
the general aggregation described below.
\end{enumerate}

The backward suffix pass costs $O(km)$ operations per row.
Each recursive insertion, merge and lexicographic comparison
costs $O(k)$ operations. The total cost of the local tail counts,
including list construction, is therefore $O\{wkN_k(m)\}$.

Final aggregation need not examine every retained set separately for every endpoint. Write $h_{\boldsymbol{\ell}}=p_{C_{\boldsymbol{\ell}}}^{(k)}$ for the local $p$-value of a completion. For each observed position $t$, let $u_t$ be the largest local $p$-value among smaller intersections containing $j_{(t)}$ and completions in which $j_{(t)}$ is one of the $k-1$ separately retained endpoints. For each suffix starting position $s$, let $v_s$ be the largest local $p$-value among completions whose suffix is $I_s$. With empty maxima defined as zero, set
\[
 \begin{split}
 u_t&=\max\left\{
       \max_{\substack{1\leq|C|<k\\j_{(t)}\in C}}p_C^{(k)},\quad
       \max_{\substack{1\leq\ell_1<\cdots<\ell_k\leq m\\t\in\{\ell_1,\ldots,\ell_{k-1}\}}}h_{\boldsymbol{\ell}}
       \right\},\\
 v_s&=\max_{\substack{1\leq\ell_1<\cdots<\ell_k\leq m\\\ell_k=s}}h_{\boldsymbol{\ell}},\qquad
 \widetilde p_{j_{(t)}}^{(k)}=\max\{u_t,\max_{1\leq s\leq t}v_s\}.
 \end{split}
\]
Each smaller intersection updates $u_t$ at its members' observed positions. Each completion updates $u_t$ at its $k-1$ separately retained positions and $v_{\ell_k}$ once. A forward cumulative maximum of the $v_s$ then accounts for membership in every suffix. These are exactly all ways an endpoint can belong to a retained set. Aggregation costs $O\{kN_k(m)\}$ operations. For fixed $k$, streaming permutation rows requires $O\{N_k(m)+m\}$ additional working memory for the counts, suffix lists and recursive construction, with constants depending on $k$.

For fixed $k$, $N_k(m)=O(m^k)$, so the cost after ordering and computing endpoint statistics is $O(wm^k)$; including observed sorting gives $O(m\log m+wm^k)$. The closure calculation is linear for $k=1$, quadratic for $k=2$, and cubic for $k=3$, where $N_3(m)=m+\binom m2+\binom m3$. The polynomial statement treats $k$ as fixed; for $k=m$, $N_m(m)=2^m-1$. These bounds exclude the cost of generating permutations and evaluating the endpoint statistics.

\subsection{Quadratic computation for \tea{}}\label{sec:tea-computation}
For $k=2$, write $C_{\ell_1,\ell_2}=\{j_{(\ell_1)}\}\cup I_{\ell_2}$, so $\mathcal F_2$ consists of the $m$ singletons and the $\binom m2$ completions $C_{\ell_1,\ell_2}$ with $1\leq\ell_1<\ell_2\leq m$.

For each permutation row, retain the largest two endpoint statistics in every suffix. Inserting the one value $T_{j_{(\ell_1)}}^{(r)}$ into the pair retained for suffix $I_{\ell_2}$ gives $K_{C_{\ell_1,\ell_2}}^{(r)}$ in constant time. For the final one-element suffix, insert into its single value. Count inclusive lexicographic exceedances over the $w$ evaluated permutations for every $C_{\ell_1,\ell_2}$, and scalar inclusive exceedances for every singleton.

Write $h_{\ell_1,\ell_2}=p_{C_{\ell_1,\ell_2}}^{\tea}$ for $1\leq\ell_1<\ell_2\leq m$. The general aggregation formula then becomes
\begin{equation}\label{eq:aggregation}
 \begin{split}
 u_t&=\max\{p_{\{j_{(t)}\}}^{\tea},\max_{t<s\leq m}h_{t,s}\},\\
 v_s&=\max_{1\leq\ell_1<s}h_{\ell_1,s},\qquad
 \widetilde p_{j_{(t)}}^{\tea}=\max\{u_t,\max_{1\leq s\leq t}v_s\}.
 \end{split}
\end{equation}
Here $u_t$ includes the singleton and completions in which $j_{(t)}$ is the separately retained endpoint; the cumulative maximum of $v_s$ covers completions whose suffix contains it. These are exactly the two ways an endpoint can belong to a completion $C_{\ell_1,\ell_2}$. Formula~\eqref{eq:aggregation} therefore aggregates the local $p$-values in $O(m^2)$ time, rather than scanning each set separately for each endpoint.

Computing all permutation tail counts costs $O(wm^2)$ operations after the endpoint statistics and observed order are available, giving $O(m\log m+wm^2)$ when observed sorting is included. Streaming through permutation rows uses $O(m^2)$ additional memory for the local counts and $O(m)$ for the current suffix maxima. For fixed $w$, the closure calculations after observed sorting are quadratic in $m$ for \tea{} and linear for shared-uniform maxT; neither bound includes permutation generation or endpoint-statistic evaluation.

For $m\geq2$, the local $p$-value for the full intersection is $p_{C_{1,2}}^{\tea}$. It should be distinguished from the adjusted $p$-values, since rejection of the full intersection alone does not ensure that all other intersections needed for an elementary rejection are rejected. 

\subsection{Why equality of the observed threshold matters}
For $m=4$ and observed order $1,2,3,4$, the retained nonsingleton sets are
\[
 \{1,2,3,4\},\ \{1,3,4\},\ \{1,4\},\
 \{2,3,4\},\ \{2,4\},\ \{3,4\}.
\]
The four singletons complete the family. For example, $\{1,3\}$ is covered by $\{1,3,4\}$, with the same observed first two statistics. Both their permuted statistics are compared with that common threshold, so increasing the permuted pair can only increase the tail count. For an arbitrary increasing set statistic, such as a sum of nonnegative endpoint statistics, adding an endpoint can also increase the observed threshold. Monotonicity of the permuted values then does not imply monotonicity of the tail probabilities. This is the structural reason the stated quadratic shortcut applies to the two-endpoint statistic.

\section{A finite conditional power comparison}\label{sec:example}
\subsection{The experiment and the closed decisions}
Consider two count endpoints, one treated subject and two controls, and condition on the three labelled pooled profiles
\[
 z_1=(3,0),\qquad z_2=(0,4),\qquad z_3=(0,2).
\]
Let $a\in\{1,2,3\}$ index the treated subject. Under the conditional complete null the three assignments are equally likely. At endpoint $j$, use the ordinary difference $T_j=Y_{{\rm tr},j}-\overline Y_{{\rm co},j}$. Enumerating the six subject permutations counts each assignment twice, giving the same tail proportions as enumerating the three assignments. Direct calculation yields Table~\ref{tab:example}.

\begin{table}[ht]
\centering
\caption{The finite conditional experiment: endpoint statistics and local $p$-values}
\begin{tabular}{crrrrrrr}
Treated subject & $T_1$ & $T_2$ & $K_{\{1,2\}}$ & $p_{\{1\}}^{\mathit{cl}}$ & $p_{\{2\}}^{\mathit{cl}}$ & $p_{\{1,2\}}^{\mathit{cl}}$ & $p_{\{1,2\}}^{\tea}$\\[4pt]
1 & $3$ & $-3$ & $(3,-3)$ & $1/3$ & $1$ & $2/3$ & $2/3$\\
2 & $-3/2$ & $3$ & $(3,-3/2)$ & $1$ & $1/3$ & $2/3$ & $1/3$\\
3 & $-3/2$ & $0$ & $(0,-3/2)$ & $1$ & $2/3$ & $1$ & $1$
\end{tabular}
\label{tab:example}
\end{table}

At $\alpha=1/3$, \tea{} closure rejects $H_2$ exactly at assignment 2, because both its singleton and full-intersection values equal $1/3$ there. It never rejects $H_1$. Classical closure rejects neither hypothesis in any assignment. For shared-uniform randomization, the full-intersection values at assignments 1, 2 and 3 are $2U/3$, $2U/3$ and $2/3+U/3$. The relevant singleton values at the first two assignments are $U/3$. Hence shared-uniform closure rejects $H_1$ at assignment 1 when $U\leq1/2$, rejects $H_2$ at assignment 2 when $U\leq1/2$, and has no other rejections.

\subsection{The alternative and conditional optimality}
Assume independent subjects with independent Poisson outcomes at the two endpoints. At endpoint $j$, the control mean is $\lambda_j>0$ and the treatment mean is $e^\beta\lambda_j$, for $\beta>0$. The likelihood ratio for assigning profile $z_a$ to treatment rather than control is a constant times $\exp\{\beta(z_{a1}+z_{a2})\}$. The constant cancels between assignments because the group sizes are fixed. Writing $x=e^\beta$, the conditional assignment probabilities are therefore
\[
 \pr_\beta(a=1)=\frac{x}{1+x+x^2},\qquad
 \pr_\beta(a=2)=\frac{x^2}{1+x+x^2},\qquad
 \pr_\beta(a=3)=\frac{1}{1+x+x^2}.
\]

\begin{proposition}\label{prop:finite-power}
In this conditional experiment, the event of at least one rejection by \tea{} closure is uniformly most powerful at level $1/3$ against $\beta>0$, among all possibly randomized tests of the conditional complete null. Its power and that of shared-uniform family rejection are
\[
 \pi_{\tea}(\beta)=\frac{x^2}{1+x+x^2},\qquad
 \pi_U(\beta)=\frac{x+x^2}{2(1+x+x^2)},
\]
with
\[
 \pi_{\tea}(\beta)-\pi_U(\beta)
 =\frac{x(x-1)}{2(1+x+x^2)}>0.
\]
Both have conditional complete-null family-wise error $1/3$. Classical closure has zero power at this level.
\end{proposition}
\begin{proof}
Write $r_a\in[0,1]$ for the probability of rejection at assignment $a$, integrating any auxiliary randomization. Conditional size at most $1/3$ is equivalent to $r_1+r_2+r_3\leq1$. For every $\beta>0$, assignment 2 has the unique largest alternative probability. Hence
\[
 \sum_a r_a\pr_\beta(a)
 \leq\pr_\beta(a=2)\sum_a r_a
 \leq\pr_\beta(a=2),
\]
with equality for the same test $r_2=1$, $r_1=r_3=0$ throughout the alternative family. Two-endpoint augmentation implements precisely this family rejection event. Its power is therefore $\pi_{\tea}(\beta)$. The shared-uniform decisions above give $\pi_U(\beta)$, and subtraction proves the strict inequality. At the complete null $x=1$, both powers equal $1/3$.
\end{proof}

The augmented full-intersection test has the same rejection event here. Each closed procedure makes at most one rejection, and both elementary hypotheses are false under the stated alternative, so its expected number of true discoveries equals its family rejection power. At rate ratio two these powers are $4/7$ and $3/7$. The comparison does not hold separately at every endpoint: shared-uniform closure has positive probability of rejecting $H_1$, whereas \tea{} never rejects it in this experiment. The conditional optimality claim concerns this specified experiment and alternative family, not arbitrary permutation models.

\subsection{Information within a tie}\label{sec:tie-information}
Fix $k\geq2$ and an intersection $I$ with $|I|\geq2$, and condition on its fully enumerated permutation orbit. Let $C$ be the full set of configurations on this orbit with scalar maximum $M_I=c$. The conditional null and alternative laws on $C$ are $P_{0,C}$ and $P_{1,C}$, and $\ell_C=dP_{1,C}/dP_{0,C}$, so $E_0(\ell_C\mid C)=1$. All expectations below also condition on the fixed orbit. The alternative is arbitrary, provided it assigns positive probability to $C$.

Let $a$ be the null proportion on the orbit with maximum strictly larger than $c$, and $b$ the proportion with maximum equal to $c$. These are the common values of $a_I$ and $b_I$ for configurations in $C$. For $z\in C$, write $S_I(z)=(t_{2,I}(z),\ldots,t_{d_I,I}(z))$ for the remaining ordered vector, with $d_I=\min\{k,|I|\}$. For \tea{}, it contains only the second-largest endpoint statistic. Define
\[
 R_C(z)=P_{0,C}\{S_I(Z')\lexge S_I(z)\},\qquad Z'\sim P_{0,C}.
\]
Within $C$, the local augmentation $p$-value is $p_I^{(k)}(z)=a+bR_C(z)$. Under the null, configurations are uniform on the fixed orbit and hence on $C$. Ordering the equivalence classes of $S_I$ from most to least extreme, a class of size $c_h$ has rank $(c_1+\cdots+c_h)/|C|$ and probability $c_h/|C|$. Consequently,
\[
 P_{0,C}\{R_C(Z)\leq u\}\leq u,\qquad 0\leq u\leq1.
\]
If $S_I$ distinguishes all configurations in $C$, the rank is uniform on $\{1/|C|,\ldots,1\}$; otherwise, residual ties can make the rank distribution more conservative. Here $|C|$ counts distinct configurations; equal group multiplicities give the same probabilities when group elements are counted instead.

For $a<\alpha<a+b$, randomized maxT rejects every configuration in $C$ with probability $\gamma=(\alpha-a)/b$. At $\alpha\leq a$ both rules have zero conditional rejection probability, and at $\alpha\geq a+b$ both have probability one. Thus, with $\gamma=\min\{1,\max\{0,(\alpha-a)/b\}\}$, the augmentation rule is
\[
 r_\alpha(z)=\1\{R_C(z)\leq\gamma\},\qquad
 q_\alpha=E_0(r_\alpha\mid C)\leq\gamma.
\]
Writing $F_{1,C}(u)=P_{1,C}\{R_C(Z)\leq u\}$, its conditional power is $F_{1,C}(\gamma)$, whereas randomized maxT has conditional power $\gamma$. Hence augmentation is more powerful within $C$ precisely when $F_{1,C}(\gamma)>\gamma$.

For any test function $r\in[0,1]$ with $E_0(r\mid C)=q$,
\[
 E_1(r\mid C)-q
 =E_0(r\ell_C\mid C)-E_0(r\mid C)E_0(\ell_C\mid C)
 =\operatorname{cov}_0(r,\ell_C\mid C).
\]
Applying this identity to $r_\alpha$, using $E_1(r_\alpha\mid C)=F_{1,C}(\gamma)$ and subtracting $\gamma-q_\alpha$ proves equation~(11) of the main paper. At a common conditional size $q$, the constant randomized rule has power $q$, whereas a deterministic rule has greater power precisely when the covariance is positive. At the nominal threshold $\alpha$, the \kea{} test exceeds randomized maxT within this tie precisely when this covariance exceeds the unused conditional null mass $\gamma-q_\alpha$.

Write $S=S_I(Z)$ for the remaining ordered vector. A sufficient condition for a nonnegative covariance is that
\[
 h(s)=E_0(\ell_C\mid S=s,C)
\]
is nondecreasing in lexicographic order. Indeed, $r_\alpha$ is a function of $S$ and is nondecreasing in that same order, and
\[
 \operatorname{cov}_0(r_\alpha,\ell_C\mid C)
 =\operatorname{cov}_0\{r_\alpha(S),h(S)\mid C\}
 =\tfrac12 E_0[\{r_\alpha(S)-r_\alpha(S')\}\{h(S)-h(S')\}\mid C]
 \geq0,
\]
where $S,S'$ are now independent conditional null draws. The inequality is strict if the two factors have the same nonzero sign with positive probability. At equal conditional size this gives a strict power gain; otherwise it must exceed $\gamma-q_\alpha$. The condition states that lexicographically larger remaining vectors carry more evidence for the alternative within a tie in the maximum. For \tea{}, it reduces to the condition that larger values of the second-largest endpoint statistic carry more evidence within that tie.

The comparison concerns a single intersection, so a local power gain need not translate into an elementary rejection after closure. The finite experiment in Sections~S4.1--S4.2 shows that the gain can survive closure.

\section{Simulation specifications}\label{sec:simulations}
\subsection{Data-generating mechanism and statistic}
The simulations use binary endpoints and independent subjects in two balanced groups. Total sample sizes are $n=n_0+n_1\in\{20,50,100,250\}$, with $n_0=n_1$, and family sizes are $m\in\{10,100,1000\}$. In each dataset $b$, a set $I_{1b}$ of exactly $m/10$ signal endpoints is selected uniformly without replacement; its complement $I_{0b}$ contains the $9m/10$ true nulls. Signal positions are redrawn independently for each dataset. Control event probabilities are $p_{0j}=0.005$ for all endpoints. Treatment probabilities are $p_{1j}=0.005+\Delta(n,m)$ for $j\in I_{1b}$ and $p_{1j}=0.005$ otherwise, where
\begin{equation}\label{eq:effect}
 \Delta(n,m)=0.33(10/n)^{0.48}
 \bigl[1+0.15\max\{0,\log_{10}(m/10)\}\bigr].
\end{equation}
All probabilities lie strictly between zero and one on the stated grid. Both power and family-wise error are evaluated in this mixed configuration, with all $m$ endpoints entering each multiple-testing procedure.

For subject $i$ in group $G_i\in\{0,1\}$, generate
\[
 W_{ij}=\sqrt\rho F_i+\sqrt{1-\rho}\,\varepsilon_{ij},\qquad
 Y_{ij}=\1\{W_{ij}\leq\Phi^{-1}(p_{G_i j})\},
\]
where the $F_i$ and $\varepsilon_{ij}$ are mutually independent standard-normal variables, independent also of the signal selection, and $\Phi$ is the standard-normal distribution function. The factor $F_i$ is shared across endpoints within subject $i$. The main experiment uses $\rho=0$, giving independent endpoints conditional on the signal set. Section~\ref{sec:correlated} uses $\rho=0.5$ with the same margins and effect sizes. Here $\rho$ is the pairwise correlation of the latent Gaussian variables, not of the thresholded binary outcomes. Effect sizes vary across design cells according to \eqref{eq:effect}; all method comparisons are within the same cell.

Let $X_{gj}=\sum_{i:G_i=g}Y_{ij}$ be the group success count, $\widehat p_{gj}=X_{gj}/n_g$, and $\widehat p_j=(X_{0j}+X_{1j})/n$. The endpoint statistic is
\begin{equation}\label{eq:binary-statistic}
 T_j=\frac{\widehat p_{1j}-\widehat p_{0j}}
 {\{\widehat p_j(1-\widehat p_j)(n_0^{-1}+n_1^{-1})\}^{1/2}},
\end{equation}
with value zero when the denominator vanishes. It is directed towards increased treatment incidence. Degenerate endpoints are retained. Complete outcome rows are permuted across fixed group positions, applying the same permutation to every endpoint. The pooled proportion is invariant under these permutations. For either value of $\rho$, conditional on the signal set, the true-null outcome vectors have the same joint distribution in both groups and are exchangeable across group positions, including when other endpoints are signals. The intersection-null invariance assumption therefore holds in both designs.

\subsection{Methods and auxiliary variables}
Every design cell contains 1000 datasets, each with 10\% signals and 90\% true nulls. Each permutation analysis uses 1000 independently sampled permutations plus the identity, so $w=1001$, with the identity at $r=1$. All five methods in the independent and correlated experiments use the same datasets within each cell; the three closed maxT procedures also use the same permutation lists. Classical maxT and shared-uniform maxT use their linear suffix shortcuts. The \tea{} procedure is \kea{} with $k=2$ and uses the quadratic family and aggregation in Section~\ref{sec:tea-computation}, with classical singleton tests. These are exact evaluations of full closure, not approximations obtained by dropping intersections.

Section~\ref{sec:higher-k} compares \tea{} with $k=3,4$ using the independent design and omitting $m=10$.

One independent auxiliary uniform is generated for each dataset in the shared-uniform procedure and reused across all intersections. It is not redrawn for each permutation or endpoint. Its realized value is used to make all closed-testing decisions. The randomized marginal comparator uses independent endpoint uniforms, independent also of the data and permutations. Separate random-number streams for data, permutation sampling and auxiliary draws allow the latter to be varied or recorded without changing the dataset or its permutation distribution.

For the randomized marginal comparator, write $S_j=X_{0j}+X_{1j}$ for the pooled success count and $x_j=X_{1j}$ for the observed treated count. Under its conditional null, $X_{1j}$ has the hypergeometric law
\[
 f_j(x)=\frac{\binom{S_j}{x}\binom{n-S_j}{n_1-x}}{\binom n{n_1}},
\]
on its feasible integer support. The randomized marginal value is
\[
 p_j^{\rm marg}=\sum_{x>x_j}f_j(x)+V_jf_j(x_j),
 \qquad V_1,\ldots,V_m\stackrel{\rm iid}{\sim}\Unif(0,1).
\]
Holm adjustment is then applied \citep{holm1979,habigerpena2011}. If $p_{(1)}^{\rm marg}\leq\cdots\leq p_{(m)}^{\rm marg}$ are the ordered marginal values, their adjusted $p$-values are
\[
 \widetilde p_{(k)}^{\rm Holm}
 =\min\{1,\max_{\ell\leq k}(m-\ell+1)p_{(\ell)}^{\rm marg}\}.
\]
This order is the marginal $p$-value order, not necessarily the observed endpoint-statistic order used for the maxT shortcuts. Hypergeometric probabilities remove permutation-sampling resolution as a limitation for this comparator. A degenerate marginal distribution gives $p_j^{\rm marg}=V_j$.

The asymptotic comparator applies the same Holm adjustment to $1-\Phi(T_j)$ for nondegenerate endpoints and to one for degenerate endpoints. Holm does not require independent marginal $p$-values for strong error control when they are valid. Nevertheless, its Bonferroni local tests differ from permutation maximum tests. Comparing randomized marginal Holm with shared-uniform maxT therefore changes the local tests and the multiplicity calculation; it does not isolate the effect of coupling intersection uniforms.

\subsection{Power, family-wise error and Monte Carlo uncertainty}
At level $\alpha=0.05$, let $R_{bj}$ indicate rejection of endpoint $j$ in dataset $b$, with signal set $I_{1b}$ and true-null set $I_{0b}$ as defined above. For $N=1000$ datasets, the estimated average endpoint power is
\[
 \widehat\pi=\frac1N\sum_{b=1}^N A_b,\qquad
 A_b=\frac1{|I_{1b}|}\sum_{j\in I_{1b}}R_{bj}.
\]
Its Monte Carlo standard error is $s_A/N^{1/2}$, where $s_A$ is the sample standard deviation of the dataset-level fractions $A_b$. Intervals use this standard error and the 0.975 quantile of a Student distribution with $N-1$ degrees of freedom, clipped to $[0,1]$. This treats the dataset, rather than the endpoint, as the independent replication unit and accommodates dependence among decisions induced by closure and the shared uniform.

For the same datasets, define $F_b=\1\{\sum_{j\in I_{0b}}R_{bj}>0\}$. The estimated family-wise error is $\widehat f=N^{-1}\sum_b F_b$, with a 95\% Wilson interval. This is the probability of at least one false rejection among the 90\% true nulls under a partial-null configuration, often termed partial-null FWER; it is the usual family-wise error criterion, with all $m$ hypotheses tested. Specifically, with $z=\Phi^{-1}(0.975)$, its centre and half-width are
\[
 \frac{\widehat f+z^2/(2N)}{1+z^2/N},\qquad
 \frac{z}{1+z^2/N}
 \left\{\frac{\widehat f(1-\widehat f)}{N}+\frac{z^2}{4N^2}\right\}^{1/2}.
\]
The intervals quantify simulation uncertainty and are pointwise, not simultaneous across panels or methods. At a true error probability of 0.05, the binomial Monte Carlo standard error from 1000 datasets is approximately 0.0069. Paired power differences use the dataset-level differences between the two rejection fractions, with the same Student-based standard error calculation. Their intervals are expressed in percentage points and are not clipped to $[0,1]$.

\subsection{Correlated endpoints}\label{sec:correlated}
\begin{figure}[t]
\centering
\includegraphics[width=\linewidth]{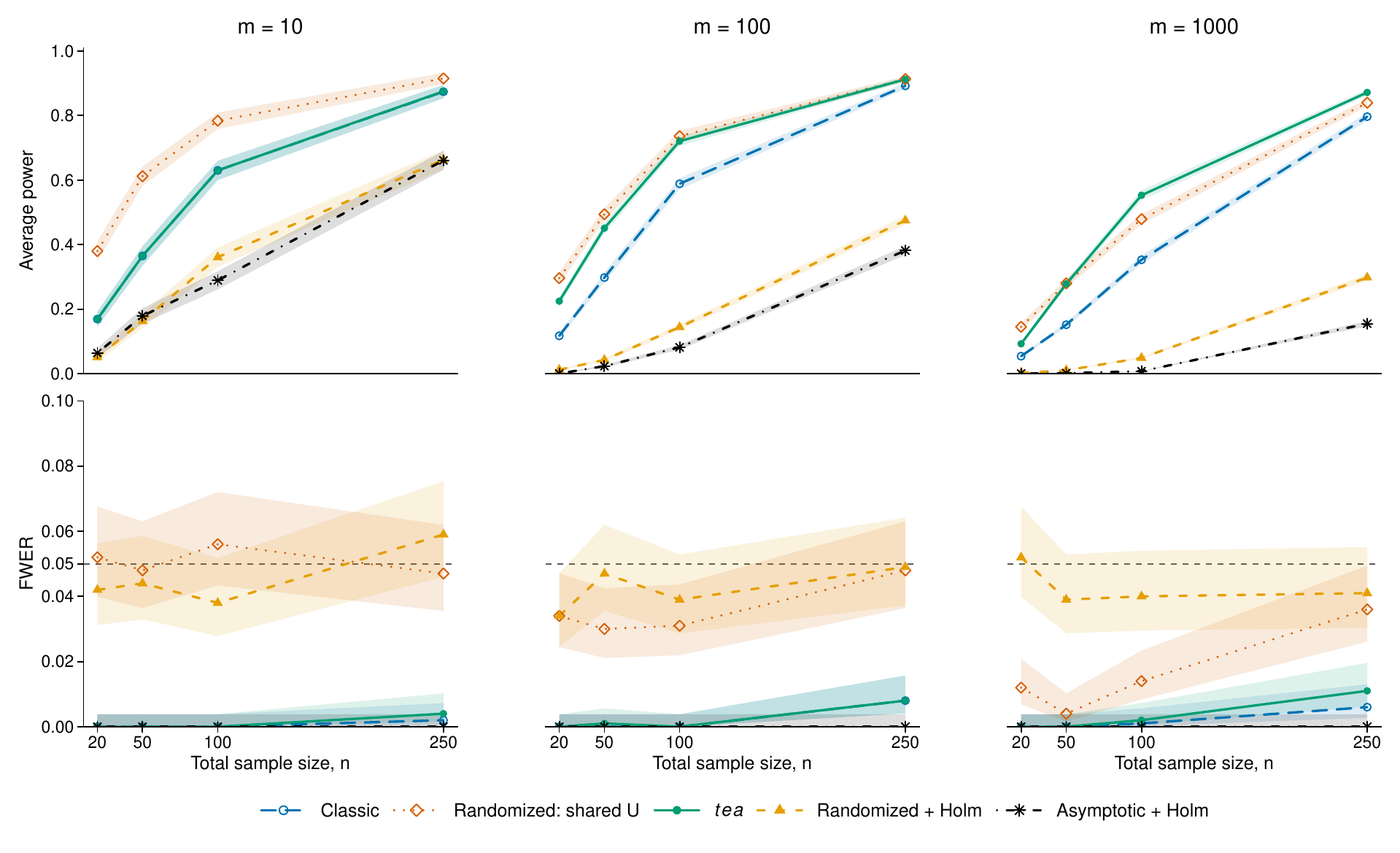}
\caption{The five procedures from the main experiment with correlated binary endpoints, generated using latent Gaussian correlation $\rho=0.5$. Every dataset has 10\% signal endpoints; each scenario uses 1000 datasets and 1000 sampled permutations plus the identity. Upper panels show average power over signals; lower panels show family-wise error over the 90\% true nulls in the same datasets. Blue circles, Classic; vermilion diamonds, shared uniform; green filled circles, \tea{}; orange triangles, randomized with Holm; black asterisks, asymptotic with Holm. Bands are 95\% Monte Carlo intervals; horizontal dashed lines mark 0.05.}
\label{fig:power-fwer-correlated}
\alttext{With correlated endpoints, shared-uniform randomization exceeds two-endpoint augmentation for 10 or 100 endpoints. With 1000 endpoints, augmentation exceeds shared-uniform randomization at sample sizes 100 and 250, is nearly equal at 50, and is less powerful at 20. Both improve on classical maxT with many endpoints. Augmentation remains conservative for family-wise error among the true nulls.}
\end{figure}

Figure~\ref{fig:power-fwer-correlated} repeats the main experiment with $\rho=0.5$ in the Gaussian construction above. Sample sizes, family sizes, the 10\% signal fraction, event probabilities, effect sizes, number of replications and permutation scheme are unchanged. The binary correlations depend on the margins: the correlation between two true-null endpoints is approximately 0.095, whereas that between two signal endpoints in the treatment group ranges from 0.239 to 0.320 across cells with at least two signals. Thus 0.5 describes the latent dependence, rather than a common correlation of the observed binary endpoints.

Both shared-uniform randomization and \tea{} improve on classical maxT when many endpoints are tested. With $m=1000$ and $n=100$, their estimated powers are 47.9\% and 55.3\%, respectively, compared with 35.3\% for classical maxT. At $n=250$, \tea{} also exceeds shared-uniform randomization, with powers 87.2\% and 84.0\%. The ordering reverses at $n=20$, where shared-uniform randomization gives 14.5\% and \tea{} 9.2\%; at $n=50$ their powers are nearly equal. With 10 or 100 endpoints, shared-uniform randomization remains more powerful in the displayed cells. The estimated family-wise error of \tea{} is at most 0.011. These results extend the comparison to dependent endpoints and reinforce that the relative power of augmentation and shared-uniform randomization depends on the alternative.

\FloatBarrier
\subsection{Higher-order endpoint augmentation}\label{sec:higher-k}
\begin{figure}[t]
\centering
\includegraphics[width=\linewidth]{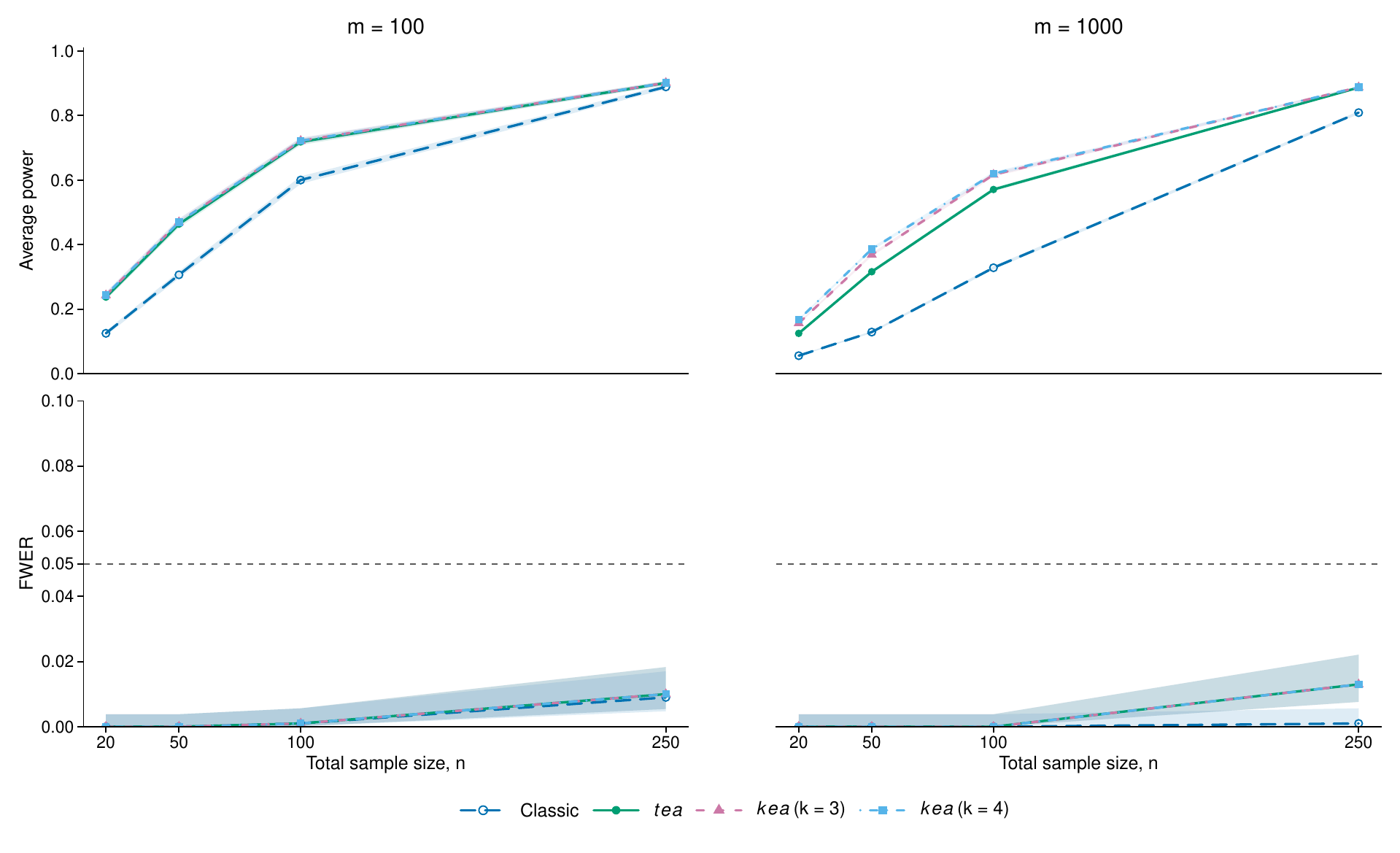}
\caption{Classical closed maxT, \tea{} and \kea{} with $k=3,4$ under the independent-endpoint design of the main experiment, restricted to $m=100$ and $m=1000$. The same datasets and permutation lists are used, with 10\% signals, 1000 datasets per scenario and 1000 sampled permutations plus the identity. Upper panels show average power over signals; lower panels show family-wise error over the 90\% true nulls in the same datasets. Blue circles, Classic; green filled circles, \tea{}; purple triangles, \kea{} with $k=3$; sky-blue squares, \kea{} with $k=4$. Bands are 95\% Monte Carlo intervals; horizontal dashed lines mark 0.05.}
\label{fig:power-fwer-k}
\alttext{Higher-order augmentation gives small additional gains with 100 endpoints. With 1000 endpoints, the third component improves power more visibly at sample sizes 20, 50 and 100; the fourth gives smaller further gains. At sample size 250 the gains beyond two components are minimal. The three augmentation procedures have the same estimated family-wise error in every displayed cell.}
\end{figure}

Figure~\ref{fig:power-fwer-k} compares classical maxT, \tea{} and \kea{} with $k=3,4$. It uses exactly the independent-endpoint design of the main experiment, retaining $m\in\{100,1000\}$ and all four sample sizes. The signal fraction, effect sizes, replications, datasets and permutation lists are unchanged. In particular, the classical and \tea{} results coincide with the corresponding main-paper results. Shared-uniform randomization and the two Holm comparators are omitted from this comparison.

The simulations compute exact closed-testing decisions at $\alpha=0.05$, rather than all adjusted $p$-values. Computational bounds are used to skip calculations that cannot affect these decisions.

With $m=100$, the gain of $k=4$ over \tea{} is at most 0.64 percentage points. With $m=1000$, higher-order augmentation gives clearer gains, mainly through the third component. At $n=50$, estimated powers are 31.6\%, 36.8\% and 38.6\% for $k=2,3,4$, respectively. The paired gain from $k=2$ to $k=3$ is 5.26 percentage points (95\% Monte Carlo interval 4.77 to 5.74), and that from $k=3$ to $k=4$ is 1.76 points (1.48 to 2.05). At $n=100$, the corresponding gains are 4.58 and 0.32 points; at $n=250$, both are below 0.1 points. The estimated family-wise error is identical for $k=2,3,4$ in every displayed cell and is at most 0.013.

These results support \tea{} as a practical compromise between power and its quadratic shortcut cost, while showing that increasing $k$ can yield further power with many endpoints. The generally smaller gain from the fourth component illustrates diminishing returns in this design. Choosing a larger fixed $k$ trades these potential gains against the higher-order computational cost.

\FloatBarrier